\documentclass[epj]{webofc}
\usepackage[utf8]{inputenc}
\usepackage[varg]{txfonts}   
\usepackage{booktabs}
\usepackage{xcolor}
\definecolor{darkred}{rgb}{0.4,0.0,0.0}
\definecolor{darkgreen}{rgb}{0.0,0.4,0.0}
\definecolor{darkblue}{rgb}{0.0,0.0,0.4}
\usepackage[bookmarks,linktocpage,colorlinks,
    linkcolor = darkred,
    urlcolor  = darkblue,
    citecolor = darkgreen]{hyperref}
\usepackage{subfigure}
\usepackage{braket}
\usepackage{latexsym}
\usepackage{epstopdf}
\usepackage{grffile}
\wocname{EPJ Web of Conferences}
\woctitle{Lattice2017}
\newcommand\nc{N_\mathrm{c}}

\newcommand\mur{\mu_\mathrm{R}}
\newcommand\mui{\mu_\mathrm{I}}

\newcommand\qrw{\theta_\mathrm{RW}}
\usepackage[normalem]{ulem}  

\newcommand{\comment}[1]{}

\renewcommand\sout{\bgroup \color{red} \ULdepth=-.5ex \ULset}
\begin{document}
%
\selectlanguage{english}
\title{%
Dirac-mode analysis for quark number density and its application for deconfinement transition
}
\author{%
\firstname{Takahiro M.} \lastname{Doi}\inst{1}\fnsep\thanks{Speaker. 
T.M.D. is supported by the Grantin-Aid for JSPS fellows (No.15J02108) 
and the RIKEN Special Postdoctoral Researchers Program. 
\email{takahiro.doi.gj@riken.jp} 
} \and
\firstname{Kouji} \lastname{Kashiwa}\inst{2}
}
\institute{%
Theoretical Research Division, Nishina Center, RIKEN, Wako 351-0198, Japan
\and
Yukawa Institute for Theoretical Physics,
Kyoto University, Kyoto 606-8502, Japan
}
\abstract{%
The quark number density at finite imaginary chemical potential 
is investigated in the lattice QCD using the Dirac-mode expansion. 
We find the analytical formula of the quark number density 
in terms of the Polyakov loop in the large quark mass regime. 
On the other hand, 
in the small quark mass region, 
the quark number density is investigated 
by using the quenched lattice QCD simulation. 
The quark number density is found to strongly depend on 
the low-lying Dirac modes while its sign does not change. 
This result leads to that the quark number holonomy 
is not sensitive to the low-lying Dirac modes. 
We discuss the confinement-deconfinement transition 
from the property of the quark number density and the quark number holonomy.
}
\maketitle
\def\slash#1{\not\!#1}
\def\slashb#1{\not\!\!#1}
\def\slashbb#1{\not\!\!\!#1}
\section{Introduction}\label{intro}

One of the important task in nuclear and particle physics is 
the nonperturbative understanding of 
the phase structure of Quantum Chromodynamics (QCD) at
finite temperature ($T$) and real chemical potential $(\mur)$. 
However, the confinement-deconfinement transition 
has not been fully understood yet. 
For example, we cannot find any order parameter of the deconfinement transition
in the case with dynamical quarks

The chiral and confinement-deconfinement transitions are key phenomena
for this purpose, but the confinement-deconfinement transition is not
yet fully understood comparing with the chiral transition.
Although the chiral transition can be described by the
spontaneous breaking of the chiral symmetry, but we can not
find any classical order-parameters of the confinement-deconfinement
transition in the presence of dynamical quarks. 

The Polyakov loop, which respect the gauge-invariant holonomy, 
is the exact order-parameter of the confinement-deconfinement 
transition in the infinite quark mass limit.
However, it is no longer the order parameter in the presence of dynamical quarks. 
Although other candidates for the order parameter 
of the deconfinement transition have been proposed
~\cite{Bilgici:2008qy,Fischer:2009wc,Kashiwa:2009ki,
Benic:2013zaa,Lo:2013hla,Doi:2015rsa}, 
they are also not the exact order parameter. 
Therefore, we need some extension of ordinary determinations to clearly 
discuss and investigate the confinement-deconfinement transition in the 
system with dynamical quarks. 

Recently, in Refs.~\cite{Kashiwa:2015tna,Kashiwa:2016vrl,Kashiwa:2017yvy}, 
it is found that 
the topological change of QCD thermodynamics at finite $\mui$ 
can be used to determine the confinement-deconfinement transition, 
and based on the non-trivial free-energy degeneracy, the quark number holonomy 
which is defined by the contour integral of the quark-number susceptibility of
$\theta=0 \sim 2\pi$ has been proposed as the quantum order-parameter for the
confinement-deconfinement transition. 
This argument is based on the analogy of the topological order 
discussed in Refs.~\cite{Wen:1989iv} and QCD at $T=0$~\cite{Sato:2007xc}
in the condensed matter physics.
The quark number holonomy counts the gapped points of the quark number density 
along $\theta$. 
As a results, 
it becomes non-zero/zero in the deconfined/confined phase. 
In particular, 
the quark number density at $\theta=\pi/3$ is important 
because it gives the property of the quark number holonomy. 
In order to investigate the quark number density at $\theta=\pi/3$, 
we use the Dirac-mode expansion \cite{Gongyo:2012vx}. 
And we will see the behavior of the quark number density 
by removal of the low-lying Dirac modes. 

This paper is organized as follows. 
In the section 2, we show the heavy quark mass 
expansion of the quark number density. 
In the section 3, we discuss the Dirac-mode expansion of the quark number density 
in both large and small quark mass regime. 
Section 4 is devoted to summary and discussions.

\section{large-mass expansion of quark number density}
\label{Sec:HQME}

In this study, we consider the ${\rm SU}(N_{\rm c})$ lattice QCD
on the standard square lattice. 
We denote each sites as $x=(x_1,x_2,x_3,x_4) \ (x_\nu=1,2,\cdots,N_\nu)$
and link-variables as $U_\nu(x)$.
We impose the temporal periodic boundary condition for link-variables to
generate configurations in the quenched calculation to manifest the
imaginary-time formalism.

On the lattice, the quark number density is defined as
\begin{align}
n_q
 =\frac{1}{V}\sum_x
   \Bigl\langle \bar{q}(x) \frac{\partial D}{\partial \mu}q(x) \Bigr\rangle
 =\frac{1}{V}\left\langle {\rm Tr}_{\gamma, {\rm c}}
   \left[ \frac{\partial D}{\partial \mu} \frac{1}{D+m} \right] \right\rangle,
\label{DefQuarkNumber}
\end{align}
where ${\rm Tr}_{c,\gamma}\equiv \sum_x {\rm tr}_{\gamma, {\rm c}}$ denotes the functional trace and ${\rm tr}_{\gamma, {\rm c}}$ is taken over spinor and color indices.
The Dirac operator $D$ in this article is taken as the Wilson-Dirac operator 
with quark mass $m$ and the chemical potential $\mu$ in the lattice unit as 
\begin{align}
D
=-\frac{1}{2}\sum_{k=1}^3
\left[P(+k)\hat{U}_k+P(-k)\hat{U}_{-k}\right] -\frac{1}{2}\left[{\mathrm e}^\mu P(+4)\hat{U}_4 +
{\mathrm e}^{-\mu} P(-4)\hat{U}_{-4}\right] +4\cdot \hat{1}
, \label{WilsonDiracOp}
\end{align}
where $\hat{1}$ is the identity matrix. 
The link-variable operator $\hat{U}_{\pm\nu}$ is defined by the matrix element 
\begin{align}
\langle
x | \hat{U}_{\pm\nu} |x' \rangle=U_{\pm\nu}(x)\delta_{x\pm\hat{\nu},x'},
 \label{LinkOp}
\end{align}
with $U_\nu\in {\rm SU}(N_{\rm c})$ and $P(\pm\nu)=1\mp\gamma_\nu$ with
$\nu=1,\cdots,4$. 
The chemical potential $\mu$ is the dimensionless on the lattice 
and we define $\theta\equiv\mathrm{Im}(\mu)N_\tau$.

We impose the temporal anti-periodicity and spatial periodicity for $D$. 
To that end, we add a minus sign to the matrix element of
the temporal link-variable operator $\hat U_{\pm 4}$
at the temporal boundary of $x_4=N_4(=0)$:
\begin{align}
\langle {\bf x}, N_4|\hat U_4| {\bf x}, 1 \rangle
=-U_4({\bf x}, N_4),\ \ \ \ \ \ \ 
\langle {\bf x}, 1|\hat U_{-4}| {\bf x}, N_4 \rangle
=-U_{-4}({\bf x}, 1)=-U_4^\dagger({\bf x}, N_t).
\label{eq:LVthermal}
\end{align}
In this notation, the Polyakov loop is expressed as
\begin{align}
L\equiv\frac{1}{N_c V}
 \sum_x {\rm tr}_c
 \Bigl\{\prod_{n=0}^{N_4-1} U_4(x+n\hat{4}) \Bigr\} 
=-\frac{1}{N_{\rm c}V}{\rm Tr}_c \{\hat U_4^{N_4}\}.
\label{PolyakovOp}
\end{align}
The minus sign stems from the additional minus on $U_4({\bf s}, N_t)$
in Eq.(\ref{eq:LVthermal}).

\begin{figure}[h]
\begin{center}
\includegraphics[scale=0.14]{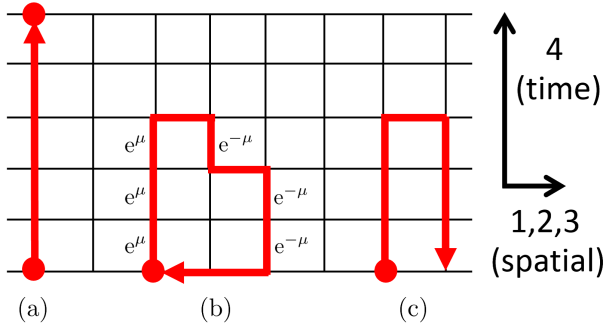}
\caption{
Several paths on the lattice in the case of $N_4=5$.
(a) A temporally closed loop. It is gauge-invariant and $\mu$-dependent.
(b) A spatially closed loop. It is gauge-invariant and $\mu$-independent.
(c) A non-closed loop. It is gauge-variant and thus it vanishes.
}
\label{loops}
\end{center}
\end{figure}

In the heavy quark mass region,
the quark number density (\ref{DefQuarkNumber}) can be expressed by
using the quark mass expansion as
\begin{align}
n_q
=\frac{1}{MV}\left\langle {\rm Tr}_{\gamma, {\rm c}} \left[ \frac{\partial D}{\partial \mu}
\sum_{n=0}^\infty\left(-\frac{\hat{D}}{M}\right)^{n} \right] \right\rangle 
\equiv\frac{1}{MV}\sum_{n=0}^\infty \frac{c^{(n)}}{(-M)^n},
\label{QuarkNumberHeavyMass}
\end{align}
where
we define the effective mass $M\equiv m+4$ and the operator $\hat{D}\equiv D-4$. 
In the case of the large quark mass, 
$c^{(n)}$ with smaller $n$ are dominant. 
The $n$-th order contribution $c^{(n)}$ has the terms 
\begin{align}
c^{(n)}=c^{(n)}_1+c^{(n)}_2+c^{(n)}_3+\cdots, 
\label{n-thContribution}
\end{align}
and $c^{(n)}_i$ is a product of $(n+1)$ link-variables.  
The examples of the paths of the products are shown in Fig.~\ref{loops}.
Note that many of them become exactly zero
because of Elitzur's theorem \cite{Elitzur:1975im};
only the gauge-invariant terms corresponding to closed loops are nonzero. 
Moreover, spatially closed loops which do not wind the temporal length 
are canceled out each other and have no contribution to the quark number density in total in each order $n$. 
Noting these important facts, it is confirmed that 
the $n$-th order contribution $c^{(n)}$ is constituted of 
the gauge-invariant loops with the length $(n+1)$ which winds the temporal direction. 
In particular, the nonzero leading term in the expansion (\ref{QuarkNumberHeavyMass})
is the $(n=N_4-1)$-th order term $c^{(N_4-1)}$ 
which relates to the Polyakov loop ($L$) and its complex conjugate (${\bar L}$). 
The leading term, $c^{(N_4-1)}$, to the
quark number density at finite $\theta$ is written as 
\begin{align}
c^{(N_4-1)}\sim{\mathrm e}^{i\theta}L-{\mathrm e}^{-i\theta}L^*
=2\sin(\theta+\phi)|L|. 
\label{QuarkNumberLeading}
\end{align}

In the heavy quark-mass expansion of the quark number
density~(\ref{QuarkNumberHeavyMass}), 
there are higher order terms beyond the leading terms, 
which are the Polyakov loop and its conjugate. 
By similar analysis in the case of the leading term, 
the higher-order terms can be expressed in terms of the quantities 
which correspond to loops on the lattice. 
For example, the sub-leading terms include a term proportional to the quantity 
\begin{align}
c^{(N_4+1)}_1
\equiv
{\rm Tr}_c \{\hat U_4U_1U_4U_{-1}U_4^{N_4-2}\}.
\label{c1}
\end{align}
These terms correspond to the closed paths 
which wind the temporal length and bypasses the spatial direction. 
As another example, loops winding the temporal direction twice or more 
can be contribute to the expansion (\ref{QuarkNumberHeavyMass}). 
For example, a loop winding the temporal direction twice
\begin{align}
c^{(2N_4-1)}_1
\equiv
{\rm Tr}_c \{\hat U_4^{2N_4}\}
\label{c2}
\end{align}
is a possible contribution to Eq. (\ref{QuarkNumberHeavyMass}) 
as the ($n=2N_4-1$)-th order term. 

\section{Dirac-mode expansion of the quark number density}
\label{Sec:DME}
In the following, 
we consider the quark number density in terms of the Dirac eigenmode.
In large quark mass region, 
we analytically investigate it in all order 
of the large quark mass expansion (\ref{QuarkNumberHeavyMass}). 
In small quark mass region, 
we perform the quenched lattice QCD simulation.

The Wilson-Dirac eigenvalues $\lambda_n$ are obtained from the 
eigenvalue equation as 
\begin{align}
D|n\rangle = \lambda_n|n\rangle, \label{DiracEigenEq}
\end{align}
where $|n\rangle$ is the Wilson-Dirac eigenstate.
Considering the Wilson-Dirac mode expansion of the chiral condensate, 
the low-lying eigenmodes of the operators $D$
have dominant contribution to the chiral condensate
known as Banks-Casher relation~\cite{Banks:1979yr, Giusti:2008vb}.

\subsection{large quark mass region}
We start the leading term to express it in terms of the Wilson-Dirac modes.
The leading contribution of the quark number density
in large quark mass region (\ref{QuarkNumberLeading}) is
expressed
by the Polyakov loop and its complex conjugate.
It is already known that the Polyakov loop can be expressed in terms of the
eigenmodes of the naive-Dirac operator which corresponds to the case of $r=0$
\cite{Suganuma:2014wya,Doi:2014zea}
and the Wilson-Dirac operator \cite{Suganuma:2016lnt,Suganuma:2016kva}.
In the following, we derive a different form of the Dirac spectral representation of 
the Polyakov loop using the operator $D$ on the square lattice
with the normal non-twisted periodic boundary condition for link-variables,
in both temporal and spatial directions.
We define a key quantity,
\begin{align}
I^{(N_4-1)}={\rm Tr}_{c,\gamma} (D\hat{U}_4^{N_4-1}).  \label{I0}
\end{align}
This quantity is defined as the slightly changed quantity from the Polyakov loop 
by replacing a temporal link-variable $\hat{U}_4$ 
to the Wilson-Dirac operator $D$. 
This quantity can be calculated as
\begin{align}
I^{(N_4-1)}
=2\mathrm{e}^{-\mu}N_{\rm c}VL.  \label{I0_1}
\end{align}
Thus, the quantity $I^{(N_4-1)}$ is proportional to $L$.
Here, other terms vanish because of the Elitzur's theorem 
and the trace over the Dirac indecies. 
On the other hand,
since $I^{(N_4-1)}$ in Eq. (\ref{I0}) is defined through the functional trace,
it can be expressed in the basis of Dirac eigenmodes as
\begin{align}
I^{(N_4-1)}
=\sum_n\langle n|D\hat{U}_4^{N_4-1}|n\rangle + \mathcal{O}(a) 
=\sum_n\Lambda_n \langle n|\hat{U}_4^{N_4-1}| n \rangle + \mathcal{O}(a).  \label{I0_2}
\end{align}
The $\mathcal{O}(a)$ term arises because 
Wilson-Dirac operator is not normal due to the $\mathcal{O}(a)$ Wilson term and 
the completeness of the Wilson-Dirac eigenstates has the $\mathcal{O}(a)$ error: 
\begin{align}
 \sum_n |n\rangle\langle n|=1+\mathcal{O}(a). 
 \label{Complete}
\end{align}
However, this error is controllable and can be ignored in close to the continuum limit. 

\begin{figure}[h]
\begin{center}
 \includegraphics[scale=0.15]{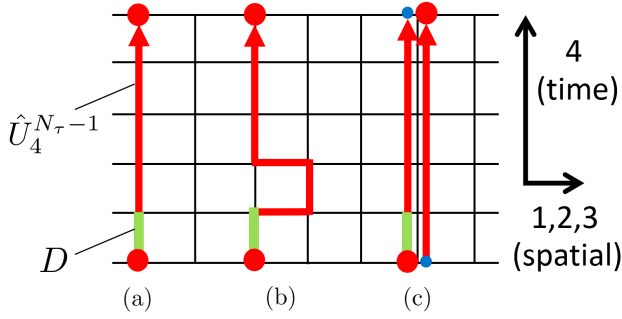}
\caption{
Schematic figures for the functional traces. 
They are defined from the gauge-invariant quantities, 
 Eqs. (\ref{PolyakovOp}), (\ref{c1}) and (\ref{c2}), 
 by changing a temporal link-variable to the Wilson-Dirac operator. 
}
\label{Fig:ECQND}
\end{center}
\end{figure}

Combining Eqs. (\ref{I0_1}) and (\ref{I0_2}), 
one can derive the relation between $L$ and the Dirac modes as
\begin{align}
L\simeq\frac{\mathrm{e}^{-\mu}}{2N_{\rm c}V}
 \sum_n\Lambda_n\langle n|\hat{U}_4^{N_4-1}| n \rangle,
 \label{RelOrig}
\end{align}
with the $\mathcal{O}(a)$ error. 
From the formula (\ref{RelOrig}), 
it is analytically found that 
the low-lying Dirac modes with $|\Lambda_n|\sim0$ have
negligible contribution to the Polyakov loop 
because the eigenvalue $\Lambda_n$ plays as the damping factor. 
It is also numerically shown that 
there is no dominant contribution in the Dirac modes 
to the Polyakov loop \cite{Doi:2014zea}. 
Thus, one can find the
Dirac spectrum representation of the quark
number density $n_q$ in the leading order 
and the low-lying Dirac modes 
have little contribution to the quark number density. 

The above discussion on the leading term can be applicable to 
the Dirac spectrum representation of the higher order terms. 
The detailed discussion is shown in our recent paper \cite{Doi:2017dyc}. 
In the same way in the case of the leading term, 
all the terms in the expansion (\ref{QuarkNumberHeavyMass}) 
can be expressed in terms of the Wilson-Dirac modes. 
Thus, it is analytically found that 
the quark number density does not depend on 
the density of the low-lying Wilson-Dirac modes in the all-order. 
However, this fact is only valid in the sufficiently large quark mass region
since other contributions which can not be expressed by $L$ and
${\bar L}$ can appear in the small $m$ region. 

\subsection{Small quark mass region}
Next, we consider the small mass regime. 
In this regime, we perform the lattice QCD simulation to investigate the quark
number density. 
In this study, we perform the quenched calculation 
with the ordinary plaquette action 
and then fermionic
observables are evaluated by using the Wilson-Dirac operator (\ref{WilsonDiracOp}) 
with the imaginary chemical potential $\mu$. 
Our calculation is performed in both the confinement phase and 
the deconfinement phase. 
In the confinement phase, 
we consider $6^4$ lattice with $\beta\equiv 6/g^2=5.6$ and $\mu=(0,1745)$ 
which corresponds to $a\simeq0.25$ fm and $T\simeq133$ MeV.
In the deconfinement phase, 
we consider $6^3\times5$ lattice with $\beta=6.0$ and $\mu=(0,2094)$ 
which corresponds to $a\simeq0.10$ fm and $T\simeq400$ MeV.
Both values of $\mu$ correspond to $\theta\simeq\pi/3$. 
In both cases, we set the quark mass as $m=-0.7$ in the lattice unit, 
which is equivalent to the hopping parameter $\kappa\equiv1/(2m+8)\simeq0.151515$, 
for the calculation of the eigenmodes of the Wilson-Dirac operator 
in the small quark mass region \cite{Aoki:1999yr}.


We calculate the quark number density as
\begin{align}
\langle n_q \rangle
=\frac{1}{2V}\left\langle {\rm Tr}_{\gamma, {\rm c}}
\left[
\frac{\partial D}{\partial \mu} \frac{1}{D+m}
-\left(\frac{\partial D}{\partial \mu} \frac{1}{D+m} \right)^\dagger
\right] \right\rangle 
\simeq\frac{i}{V}{\rm Im}\left\langle \sum_n
\Bigl\langle n \Bigl| \frac{\partial D}{\partial \mu} \Bigr|n \Bigr\rangle \frac{1}{\Lambda_n+m} \right\rangle. 
\label{QuarkNumber_Dirac}
\end{align}
This form trivially takes pure imaginary value up to the
$\mathcal{O}(a)$ error.
Each contribution, $n_q^n$, to the quark number density 
of the Dirac mode with $\Lambda_n$ 
can be defined as
\begin{align}
n_q^n=
\frac{i}{V}{\rm Im} \Bigl \langle n \Bigl| \frac{\partial D}{\partial \mu} \Bigr|n \Bigr\rangle \frac{1}{\Lambda_n+m},
\end{align}
and then the quark number density becomes
\begin{align}
 n_q=\sum_n n_q^n.
\end{align}

In Fig.~\ref{Fig:ECQND}, 
${n}_\mathrm{q}^n (\Lambda)$ is shown in the cases with
$\mu = (0,0.1745)$ and $\mu =(0,0.2094)$.
We here only show results with one particular configuration.
\begin{figure}[h]
\begin{center}
 \includegraphics[scale=0.50]{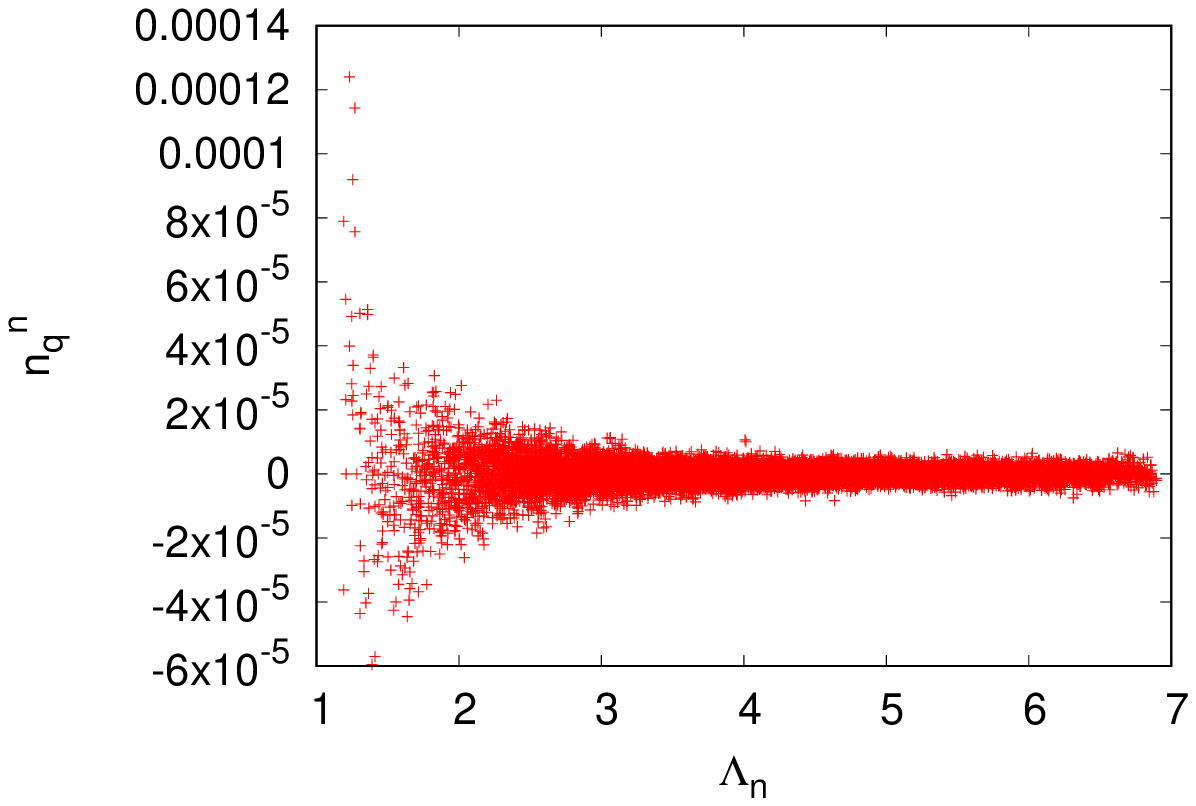}
 \includegraphics[scale=0.50]{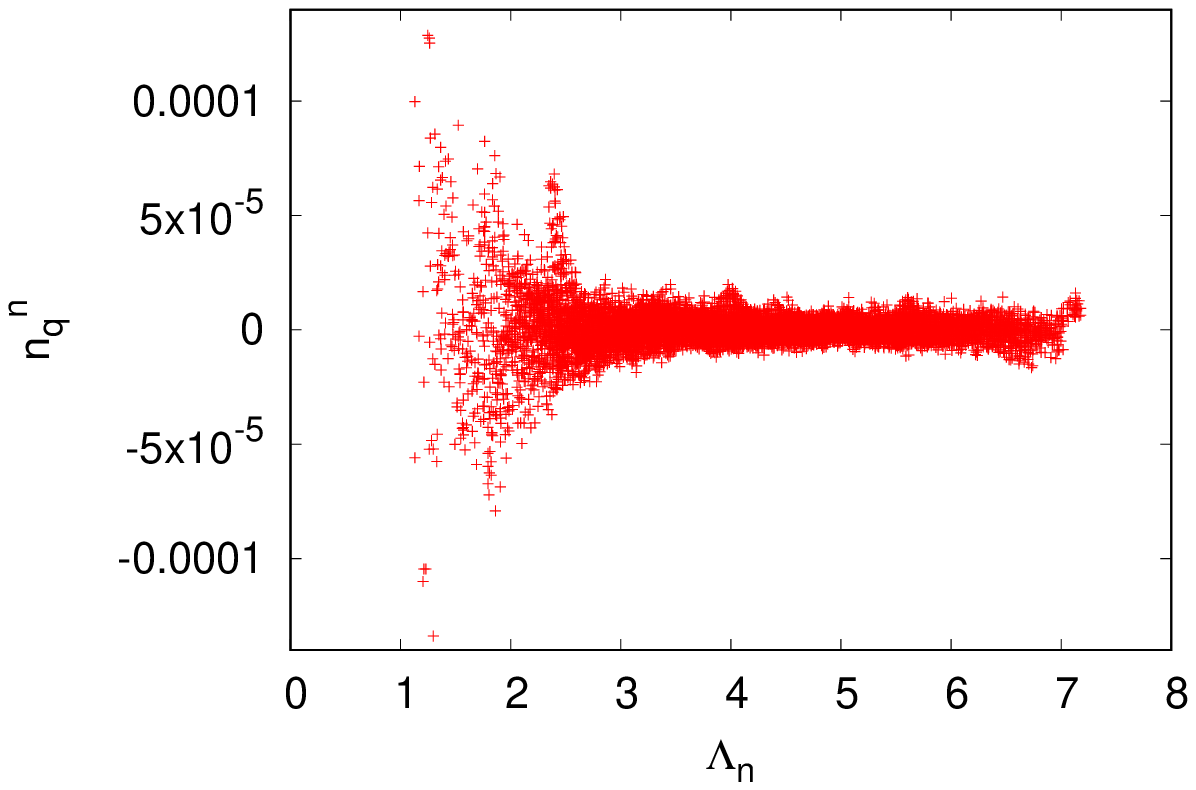}
\caption{
 Each Dirac-mode contribution ${n}_\mathrm{q}^n (\Lambda_n)$ 
 to the quark number density as a function
 plotted against the Wilson-Dirac eigenvalue $\Lambda_n$.
 Left and right panels show
 the results with $\mu =(0,0.1745)$ and $\mu =(0,0.2094)$, respectively.
 }
\label{Fig:ECQND}
\end{center}
\end{figure}

\begin{figure}[t]
\begin{center}
 \includegraphics[scale=0.50]{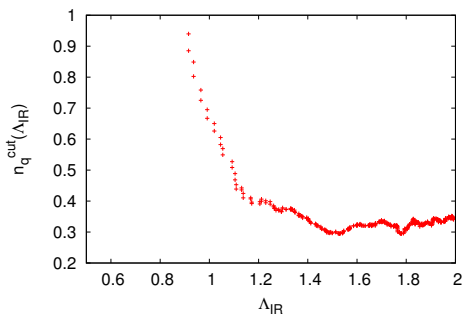}
\caption{
 The $\Lambda_\mathrm{cut}$-dependence of $n_\mathrm{q}^\mathrm{cut}$
 at $\mu =(0,0.2094)$.
}
\label{Fig:CQND}
\end{center}
\end{figure}

To investigate the quark number density in terms of Dirac modes,
we define the infra-red (IR) cutted quark number density with the cutoff
$\Lambda_\mathrm{cut}$ as
\begin{align}
 n^\mathrm{cut}_\mathrm{q} (\Lambda_\mathrm{IR})
 = \frac{1}{n_\mathrm{q}} \sum_{|\Lambda_n|>\Lambda_{\rm IR}}
 {n}_\mathrm{q}^n.
 \label{Eq:IR}
\end{align}
In Fig.~\ref{Fig:CQND}, 
we shown the result at $\mu = (0,0.2094)$, 
where the system is in the deconfinement phase. 
In general, 
we can perform the configuration average in the evaluation of Eq.~(\ref{Eq:IR}). 
However, since the averaging well works after summing over all Dirac-modes, 
it misses a physical meaning. 
Then, we show 
$n^\mathrm{cut}_\mathrm{q}$ in one particular configuration. 
Note that the sign of the quark number density 
does not changed by removing the low-lying Dirac modes 
while the absolute value of the quark number density seems to drastically change. 
This tendency has been found in almost all our configurations. 
This result means that the absolute value of the quark number density
shares a same property in terms of Dirac modes with the chiral condensate, 
while its sign shares the property with the Polyakov loop.

Finally, the topological confinement-deconfinement transition 
can be discussed in terms of the Dirac-mode analysis.
The order-parameter of the topological confinement-deconfinement 
transition can be expressed~\cite{Kashiwa:2016vrl} as 
\begin{align}
\Psi &=
   \oint_{0}^{2\pi}
   \Bigl\{\mathrm{Im} \Bigl(
          \frac{d {\tilde n}_q}{d \theta} \Bigl|_T \Bigr) \Bigr\}
         ~d\theta,
 \label{Eq:psi}
\end{align}
where ${\tilde n}$ is the dimensionless quark number density such as
${\tilde n}_\mathrm{q} = n_\mathrm{q}/T^3$.
It counts gapped points of the quark number density along 
$\theta$ direction and thus it becomes zero/non-zero 
in the confined/deconfined phase. 
The quark number holonomy (\ref{Eq:psi}) can be expressed as
\begin{align}
\Psi &= \pm
   2 \nc \lim_{\epsilon \to 0}
   \Bigl[ \mathrm{Im}~
 {\tilde n}_q ( \theta=\qrw \mp \epsilon )\Bigr],
\label{psi1}
\end{align}
when the RW endpoint which is the endpoint of the RW transition line
becomes the second-order point at
$\theta_\mathrm{RW} =\pi / 3$.
In Eq.~(\ref{psi1}), $\lim_{\epsilon \to 0} n_\mathrm{q} (\qrw \mp
\epsilon)$ characterizes
$\Psi$ and thus $\Psi$ shares the same property about the Dirac modes
with $n_\mathrm{q}$.
The important point here is that 
the absolute value of $n_q$ does not have so much meaning even if it is non-zero, 
but its sign is important since 
the sign flipping at $\theta = (2k-1) \pi/3$
characterizes the gapped points along $\theta$-direction.
From our quenched lattice QCD data, the sign of the quark number density
are insensitive to the low-lying Dirac-modes and this behavior is similar
to the Polyakov loop. 


\section{Summary and discussion}
\label{Sec:summary}

In this paper, we have discussed properties of the quark number
density at finite temperature ($T$) and imaginary chemical potential
($\mu_\mathrm{I}$) by using the Dirac-mode expansion.

From the heavy quark mass expansion with the Dirac mode expansion,
we found that low-lying Dirac modes do not dominantly contribute to the
quark number density in all order of the heavy quark mass expansion.

In the small quark mass regime, 
we found that the absolute value of the quark number density strongly
depends on low-lying Dirac modes, but its sign does not.
Our result shows that 
the quark number holonomy is sensitive to the confinement properties of QCD 
and it is the good quantum order parameter 
for the confinement-deconfinement transition.

In order to avoid the incompleteness of the Wilson-Dirac modes, 
one can consider the hermitian Wilson-Dirac operator $H\equiv\gamma_5 D$. 
In fact, the actual calculation is shown in our recent paper \cite{Doi:2017dyc}. 
Moreover, the more detailed calculation on the higher order terms in the 
large-mass expansion is also included in it. 


\bibliography{lattice2017.bib}

\end{document}